\documentclass[iop]{emulateapj}

\shorttitle{The effects of rotation and mixing in a slowly pulsating B8\,V star}
\shortauthors{P\'{a}pics et al.}

\begin{document}

\title{Asteroseismic fingerprints of rotation and mixing in\\the slowly pulsating B8\,V star KIC\,7760680}

\author{P.~I.~P\'{a}pics\altaffilmark{1}}
\affil{Instituut voor Sterrenkunde, KU Leuven, Celestijnenlaan 200D, B-3001 Leuven, Belgium}
\affil{Kavli Institute for Theoretical Physics, University of California, Santa Barbara, CA 93106, USA}
\email{Peter.Papics@ster.kuleuven.be}

\author{A.~Tkachenko\altaffilmark{1}}
\affil{Instituut voor Sterrenkunde, KU Leuven, Celestijnenlaan 200D, B-3001 Leuven, Belgium}

\author{C.~Aerts}
\affil{Instituut voor Sterrenkunde, KU Leuven, Celestijnenlaan 200D, B-3001 Leuven, Belgium}
\affil{Department of Astrophysics, IMAPP, Radboud University Nijmegen, PO Box 9010, 6500 GL Nijmegen, The Netherlands}
\affil{Kavli Institute for Theoretical Physics, University of California, Santa Barbara, CA 93106, USA}

\and

\author{T.~Van~Reeth, K.~De~Smedt, M.~Hillen, R.~{\O}stensen, E.~Moravveji\altaffilmark{2}}
\affil{Instituut voor Sterrenkunde, KU Leuven, Celestijnenlaan 200D, B-3001 Leuven, Belgium}

\altaffiltext{1}{Postdoctoral Fellow of the Fund for Scientific Research (FWO), Flanders, Belgium}
\altaffiltext{2}{Postdoctoral Fellow of the Belgian Science Policy Office (BELSPO), Belgium}

\begin{abstract}
We present the first detection of a rotationally affected series consisting of 36 consecutive high-order sectoral dipole gravity modes in a slowly pulsating B (SPB) star. The results are based on the analysis of four years of virtually uninterrupted photometric data assembled with the \textit{Kepler} Mission, and high-resolution spectra acquired using the HERMES spectrograph at the 1.2 meter Mercator Telescope. The spectroscopic measurements place KIC\,7760680 inside the SPB instability strip, near the cool edge, given its fundamental parameters of $T_\mathrm{eff}=11650\pm210\,\mathrm{K}$, $\log g=3.97\pm0.08\,\mathrm{dex}$, microturbulent velocity $\xi_\mathrm{t}=0.0^{+0.6}_{-0.0}\,\mathrm{km\,s}^{-1}$, $v \sin i=61.5\pm5.0\,\mathrm{km\,s}^{-1}$, and $[M/H]=0.14\pm0.09\,\mathrm{dex}$. The photometric analysis reveals the longest unambiguous series of gravity modes of the same degree $\ell$ with consecutive radial order $n$, that carries clear signatures of chemical mixing and rotation. With such exceptional observational constraints, this star should be considered as the \textit{Rosetta Stone} of SPBs for future modelling, and bring us a step closer to the much needed seismic calibration of stellar structure models of massive stars.
\end{abstract}



\section{Introduction}\label{intro}

Slowly pulsating B stars (SPB stars) are non-radial multi-periodic oscillators on the main sequence between B3 and B9 in spectral type, 11\,000\,K and 22\,000\,K in effective temperature, and $2.5\,\mathcal{M}_{\Sun}$ and $8\,\mathcal{M}_{\Sun}$ in mass \citep[e.g.,][Chapter 2]{2010aste.book.....A}. They pulsate in high-order gravity modes (with periods typically from 0.5 to 3\,days) which are driven by the $\kappa$-mechanism operating due to the opacity-bump associated with the iron-group elements \citep{1993MNRAS.265..588D,1993MNRAS.262..213G}. The g modes of the same degree $\ell$ with consecutive radial order $n$ are expected to be almost equally spaced in period, and deviations from this equal spacing carry information about the physical processes in the near-core region \citep{2008MNRAS.386.1487M}. This allows us to use asteroseismology to detect and interpret the fingerprints of mixing processes, such as core overshoot, diffusion, or rotationally induced mixing (originating from, e.g., internal differential rotation or meridional circulation), that influence the stellar lifetime significantly. Our understanding of these processes and their dependencies on different parameters (such as mass, rotation rates, or the presence of a magnetic field) is highly unsatisfactory to provide the much needed calibration for stellar structure and evolution models of the more massive stars with convective cores and radiative envelopes. This is partly due to the low number of in-depth studies of SPB stars, partly due to the low number (in terms of what is required for a secure forward modelling) of identified modes in the majority of these studies \citep[e.g.,][]{2015IAUS..307..154A}. Since the mentioned models are cornerstones in many fields of stellar and galactic astrophysics, these shortcomings must be remedied.

The \textit{Kepler} Mission does not only excel in exoplanet detections \citep{2010Sci...327..977B}, but it is also a goldmine for asteroseismology \citep{2010PASP..122..131G}. Thanks to the micromagnitude precision and the four year time-base of the initial mission, previously inaccessible diagnostic methods are now within reach. This led to the fist actual seismic modelling of an SPB star presented by \citet{2014A&A...570A...8P}, thanks to the unambiguous detection of 19 consecutive rotationally split $\ell=1$ gravity modes in the \textit{Kepler} photometry of the slowly rotating star KIC\,10526294. The detection of such period series combined with the approximate knowledge of fundamental parameters eliminates the time-consuming and resource-intensive need of individual mode identification from follow-up multi-colour photometry and/or line-profile variation studies \citep[e.g.,][]{2005A&A...432.1013D}.

KIC\,7760680 ($m_V=10.3$) was selected as a \textit{Kepler} GO Program target together with KIC\,10526294 and 6 other SPBs \citep[the procedure is described in][]{2013A&A...553A.127P}. In Section\,\ref{spectroscopy} we show that the star is very close to KIC\,10526294 in terms of the fundamental parameters, but it is rotating faster, which gives us the opportunity to study the effects of rotation. In Section\,\ref{photometry} the results from the \textit{Kepler} photometry are presented, specifically the detection of a rotationally influenced long period series. In Section\,\ref{conclusions} we discuss the implications and future prospects of this detection.


\section{Spectroscopy}\label{spectroscopy}
To confirm its initial classification, KIC\,7760680 was observed using the \textsc{Hermes} spectrograph \citep{2011A&A...526A..69R} installed on the 1.2 metre Mercator telescope on La Palma (Spain). A total of 8 spectra were taken: four consecutive exposures of $1000-1800$\,seconds in 2010, one 900\,s exposure in 2012, and three 1800\,s exposures -- each separated by two days -- in 2014. The resolving power in the used HRF (high resolution fibre) mode was $R \approx 85\,000$, while the achieved signal-to-noise ratio (S/N) per exposure was typically $\sim30$ in 2010 and 2012, and $\sim60$ in 2014.

The raw exposures were reduced using the dedicated HERMES pipeline, and we worked with the extracted cosmic-removed, log-resampled (which includes barycentric correction), merged spectra afterwards. As a last reduction step, all spectra were rectified following the process described by \citet{2013A&A...553A.127P}. We did not find any significant radial velocity variations in the data up to a level of $\sim5$\,km\,s$^{-1}$ that could indicate binarity, thus we immediately proceeded with calculating an average spectrum (using the inverse of the variance as weights). This mean spectrum with a S/N of $\sim120$ was used to determine the fundamental parameters.

We used the GSSP code package \citep{2011A&A...526A.124L,2012MNRAS.422.2960T} to determine atmospheric parameters of the star, such as the effective temperature ($T_\mathrm{eff}$), surface gravity ($\log g$), microturbulent and projected rotational velocities ($\xi_\mathrm{t}$ and $v \sin i$, respectively), and metallicity ($[M/H]$). The method is based on comparison of the observations with a grid of synthetic spectra; a $\chi^2$ merit function is used to judge on the quality of the fit. The error bars are represented by $1$-$\sigma$ confidence levels that were computed from a $\chi^2$ statistics, and include all possible correlations between the parameters and possible imperfections in continuum normalization, but exclude any other possible uncertainties such as NLTE effects \citep[e.g.,][]{2005A&A...435..669P,2012A&A...539A.143N} and/or incorrect atomic data.

The spectrum was fitted in the wavelength range between $4250$ and $5800$\,\AA. The bluer part of the spectrum was omitted due to increasing uncertainties in the continuum normalization and decreasing S/N of the data, whereas the red part was excluded due to contamination by telluric lines. After the first run of the code, it already became clear that the fit is insensitive to the variations of microturbulent velocity: our best fit suggested $\xi_\mathrm{t}$ of $0.0\,\mathrm{km\,s}^{-1}$ with a $1$-$\sigma$ uncertainty level of $2.0\,\mathrm{km\,s}^{-1}$. Moreover, we found the microturbulent velocity to correlate strongly with the metallicity of the star, which varied between the solar value (0.0\,dex) and 0.15\,dex, for the $\xi_\mathrm{t}$ range from $0.0$ to $2.0\,\mathrm{km\,s}^{-1}$. Fortunately, He lines are particularly sensitive to the variations of microturbulent velocity in B-type stars, and give little to no contribution to the overall stellar metallicity. Thus, we used helium lines to determine the microturbulent velocity parameter to be $0.0^{+0.6}_{-0.0}\,\mathrm{km\,s}^{-1}$, and fixed it in the next step while fine-tuning the remaining four atmospheric parameters. This procedure was repeated several times until convergence was achieved; the $1$-$\sigma$ uncertainties in $\xi_\mathrm{t}$ were propagated when computing the error bars in the other four parameters to take into account possible correlations.

\begin{figure}
\includegraphics[width=\hsize]{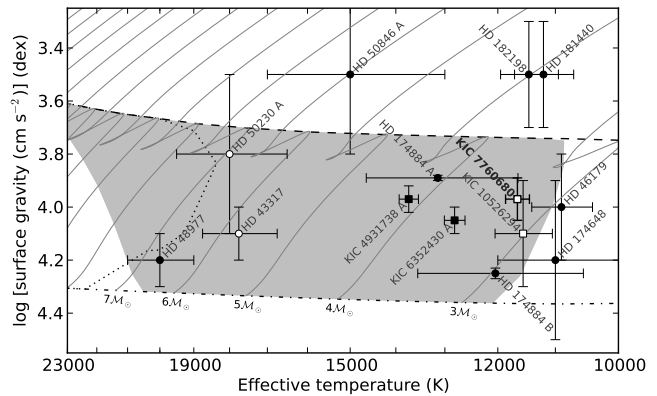}
\caption{Kiel diagram of a sample of B-type stars near the main sequence -- for which an in-depth seismic analysis was carried out -- observed by CoRoT (circles) and by \textit{Kepler} (squares). The SPBs with an observed period series are plotted using empty symbols. The dot-dashed line represents the zero-age main sequence (ZAMS), while the dashed line the terminal-age main sequence (TAMS). The thin gray lines denote evolutionary tracks for selected masses, while the SPB (gray area) and $\beta$\,Cep (cool edge plotted using a dashed line) instability strips are also shown. These were all calculated for $Z = 0.015$, $X=0.7$, using OP opacities \citep{2005MNRAS.362L...1S}, A04 heavy element mixture \citep{2005ASPC..336...25A}, and the linear nonadiabatic code by \citet{1977AcA....27...95D}, with the details of computations as in \citet{1999AcA....49..119P}. Different error bars reflect differences in data quality and methodological approach.\label{fig1}}
\end{figure}

\begin{figure}
\includegraphics[width=\hsize]{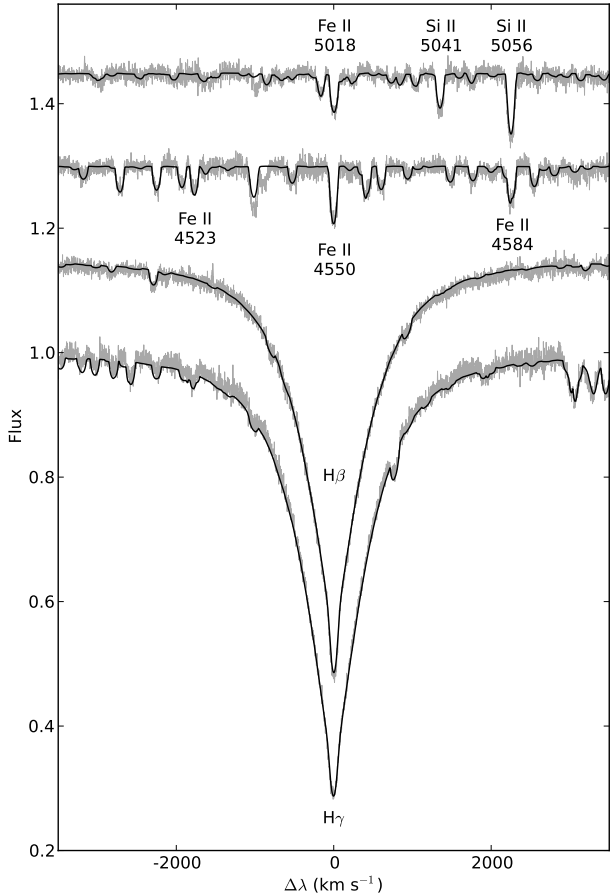}
\caption{Selected regions of the observed mean HERMES spectrum (gray) and the best fit model spectrum (black solid line). Two Balmer lines, and two wavelength ranges that are rich in metal lines are shown. A few stronger lines are labelled, and an offset of 0.15 units in flux are introduced between the selected regions for clarity.\label{fig2}}
\end{figure}

The derived parameters are $T_\mathrm{eff}=11650\pm210\,\mathrm{K}$, $\log g=3.97\pm0.08\,\mathrm{dex}$, $\xi_\mathrm{t}=0.0^{+0.6}_{-0.0}\,\mathrm{km\,s}^{-1}$, $v \sin i=62\pm5\,\mathrm{km\,s}^{-1}$, and $[M/H]=0.14\pm0.09\,\mathrm{dex}$. They place the star inside the SPB instability strip (see Figure\,\ref{fig1}) with a spectral type of B8\,V \citep[based on $T_\mathrm{eff}$ and $\log g$ values by using an interpolation in the tables given by][]{1982SchmidtKalerBook}. The obtained model fit to the mean spectrum is shown on Figure\,\ref{fig2}. We point out that the relatively large uncertainty of $v \sin i$ reflects the effect of unresolved pulsational broadening due to g-modes \citep[e.g.,][]{2014A&A...569A.118A}.

\section{Photometric Observations and Frequency Analysis}\label{photometry}

KIC\,7760680 was observed for a total of 1470.5 days (from the first commissioning quarter Q0 until the end of the last science quarter Q17, $\sim4$\,years) in Long Cadence (LC) mode with a cadence of 29.43 minutes by the \textit{Kepler} satellite. We constructed light curves for each quarter from the target pixel files using custom masks. We added extra pixels with significant flux present beyond the extent of the default mask, which resulted in quarterly curves with less instrumental trends than the standard extraction \citep[for more details see, e.g.,][]{2013A&A...553A.127P}. There are no other contaminating sources in the vicinity. After correcting for the incorrectly reported times in Q0-Q14 \citep[as noted in][]{2014A&A...570A...8P}, we cleaned all quarters from clear outliers, and detrended them using a division with a second order (except for Q0, where we used a linear) polynomial fit. Then the counts were converted to ppm and quarters were merged to a continuous light curve with a duty cycle of $91.6\%$ (part of which is shown on Figure\,\ref{fig3}).

We extracted the Fourier-parameters ($A_j$ amplitudes, $f_j$ frequencies, and $\theta_j$ phases) of the pulsation modes following a standard iterative prewhitening procedure and applied a significance criterion to select the strictly significant ones (see Figure\,\ref{fig3}), in the same way as in \citet{2014A&A...570A...8P}.

\subsection{Combination Peaks}

\begin{figure*}
\centering
\includegraphics[width=17.5cm]{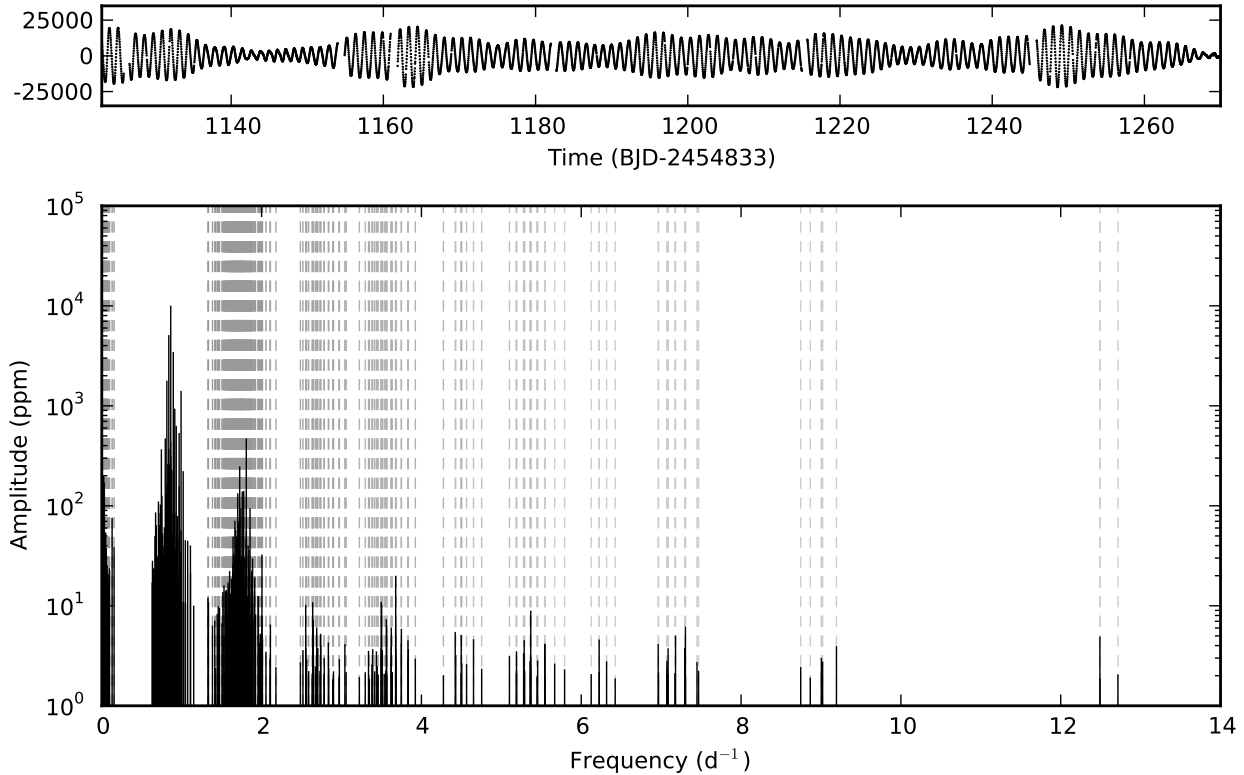}
\caption{One tenth of the reduced \textit{Kepler} light curve (top panel, brighter points towards the top), and the distribution of significant frequencies (bottom panel, black vertical lines) in the frequency spectrum of KIC\,7760680. Combination peaks are marked with gray dashed lines (darker color corresponds to simpler combinations). The region of maximum power is shown in detail on Figure\,\ref{fig4}.\label{fig3}}
\end{figure*}

The Scargle periodogram of KIC\,7760680 shown in Figure\,\ref{fig3} shows clear groups of peaks. Following the approach described by \citet{2012AN....333.1053P} we find that all peaks below $0.2\,\mathrm{d}^{-1}$ ($2.31\,\mu\mathrm{Hz}$) and above $1.2\,\mathrm{d}^{-1}$ ($13.9\,\mu\mathrm{Hz}$) are low order combination frequencies originating from independent parent modes in the range in between. Such harmonics and linear combinations are observed in many classes of classical pulsators \citep[see, e.g.,][]{2006A&A...446..237D,2011MNRAS.414.1721B,2015arXiv150402119V}, and various non-linear mechanisms may be responsible for their appeareance \citep{2011MNRAS.414.1721B}. The spectrum of independent modes is simple, and shows a clear structure. No pressure modes are observed towards shorter periods, as expected from excitation calculations for this temperature range.

\subsection{A Series of Gravity Modes}

\begin{figure*}
\centering
\includegraphics[width=17.5cm]{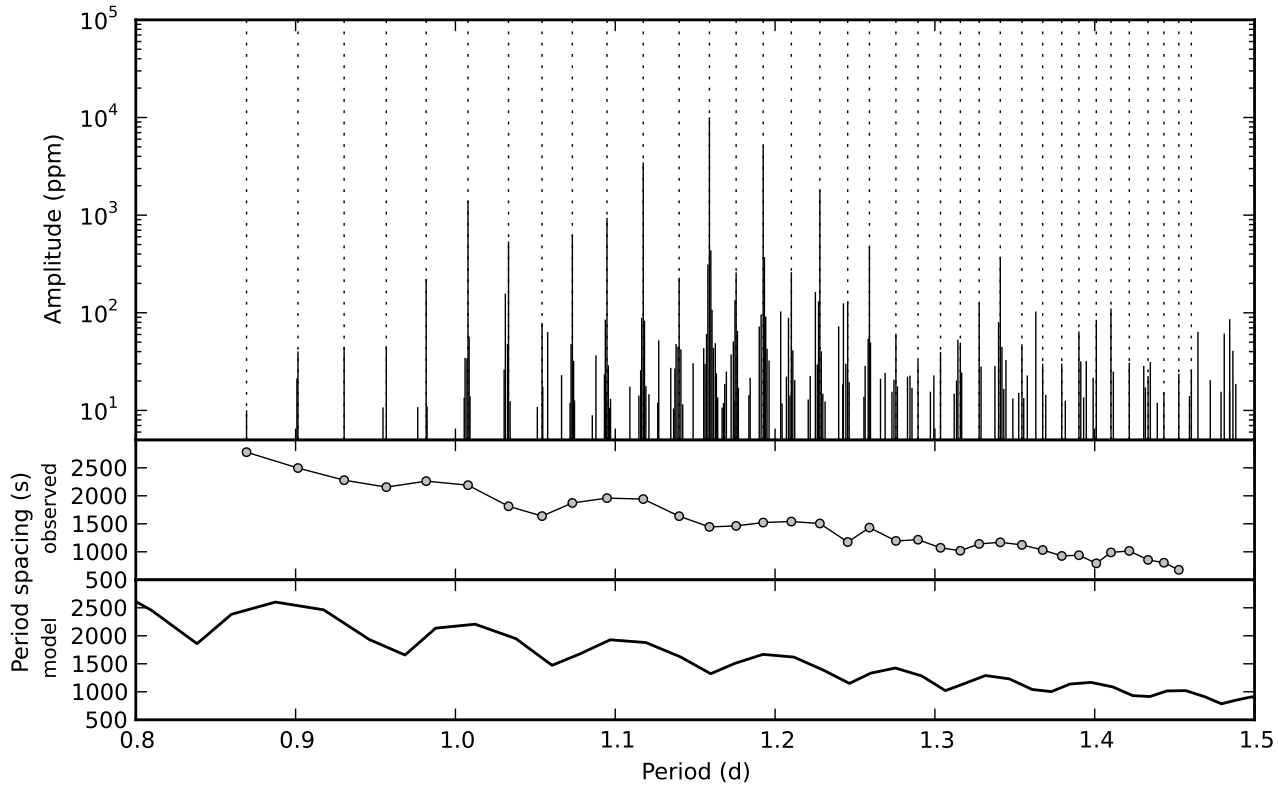}
\caption{Zoom-in of the period spectrum of KIC\,7760680 (top panel) with the members of the period spacing series marked using vertical dashed lines. The observed (middle panel) period spacing is shown along with a series from an appropriate stellar model, discussed in the text, as comparison (bottom panel). \label{fig4}}
\end{figure*}

The frequency peaks in between the mentioned groups of combination frequencies fall in the expected range for high-order gravity modes of SPB stars. Here, we detect a clear series of 36 frequencies (listed in Table\,\ref{tab1}) which follow a characteristic period spacing pattern (see Figure\,\ref{fig4}). There is a distinct cutoff in power towards longer periods at the frequency $0.63051(2)\,\mathrm{d}^{-1}$ ($7.2976(2)\,\mu\mathrm{Hz}$), which we interpret as a good proxy of the buoyancy cutoff frequency \citep[e.g.,][]{1985ApJ...297..544H,2000MNRAS.318....1T}. This cutoff value is higher than the one of KIC\,10526294 by $0.16\,\mathrm{d}^{-1}$ ($1.85\,\mu\mathrm{Hz}$), which must reflect a difference in mode energy leakage from the stellar atmosphere due to the difference in evolutionary state and/or rotation. \citet{2000MNRAS.318....1T} indeed concluded that the Coriolis force increases the cutoff frequency, although not so much for prograde sectoral modes.

This is only the fourth main sequence B-type star where we find such a series of period spacings \citep[after HD\,50230, HD\,43317, and KIC\,10526294, by][respectively]{2010Natur.464..259D,2012A&A...542A..55P,2014A&A...570A...8P}. Two of them are ultra-slow rotators, KIC\,7760680 is a moderate rotator, and HD\,43317 rotates at 50\% of its critical velocity. This new detection confirms that gravity-mode period spacings occur in pulsating B stars irrespective of their rotation rate and offers a great opportunity to probe the interior rotation profile of such stars, as first tried for KIC\,10526294 \citep{Triana2015}.

The number of consecutive peaks in the series for KIC\,7760680 is much higher than the previous record holder (19 peaks in KIC\,10526294), and this is the first time that we detect a significant tilt and a long periodic pattern in the period spacing series. There is also an apparent bifurcation in mode amplitudes around $1.2\,\mathrm{d}$, with every second peak being significantly stronger for a length of approximately 5 pairs. This implies that modes with even radial order get excited with different amplitude than modes with odd radial order near the maximum power. As Figure\,\ref{fig4} shows, this difference in amplitude is not connected to mode trapping in the inner near-core regions; it rather seems to indicate a difference in the pulsation mode energy distribution in the envelope for odd versus even radial order dipole modes.

\begin{table*}
\begin{center}
\caption{Fourier parameters (periods, frequencies, amplitudes, and phases) of the modes of the detected period series. The displayed S/N values are calculated in a window of $1\,\mathrm{d}^{-1}$ in the residual spectrum centered on the given frequency. Numbers in parentheses are the formal errors of the last significant digit.\label{tab1}}
\begin{tabular}{lccccc}
\tableline\tableline
\# & $p$ & $f$ & $A$ & $\theta$ & S/N\\
   & $\mathrm{d}$ & $\mathrm{d}^{-1}$ & $\mathrm{ppm}$ & $2\pi/\mathrm{rad}$ &  \\
\tableline
 1 &  0.86930(2)   &  1.15035(3)   &     9.9(7) &  -0.43(7)   &    4.2\\
 2 &  0.90147(1)   &  1.10930(1)   &    40(2)   &  -0.21(4)   &   13.3\\ 
 3 &  0.93036(1)   &  1.07485(1)   &    44(2)   &   0.49(4)   &    8.5\\
 4 &  0.95674(1)   &  1.04521(1)   &    45(2)   &   0.42(4)   &    8.6\\
 5 &  0.981691(6)  &  1.018650(6)  &   219(3)   &   0.06(2)   &   22.0\\
 6 &  1.007872(2)  &  0.992189(2)  &  1386(7)   &  -0.074(5)  &   60.3\\
 7 &  1.033219(4)  &  0.967849(4)  &   526(5)   &   0.28(1)   &   35.1\\
 8 &  1.05421(1)   &  0.94858(1)   &    78(2)   &   0.48(3)   &   11.4\\
 9 &  1.073164(5)  &  0.931824(4)  &   620(6)   &   0.215(9)  &   36.6\\
10 &  1.094833(4)  &  0.913381(3)  &   921(6)   &   0.399(7)  &   48.6\\
11 &  1.117507(1)  &  0.894849(1)  &  3400(11)  &   0.326(3)  &   84.8\\
12 &  1.139980(8)  &  0.877208(6)  &   225(3)   &   0.33(1)   &   21.7\\
13 &  1.158919(1)  &  0.862873(1)  &  9876(26)  &   0.409(3)  &   96.9\\
14 &  1.175637(7)  &  0.850603(5)  &   255(4)   &  -0.37(1)   &   24.2\\
15 &  1.192566(1)  &  0.838528(1)  &  5017(18)  &  -0.201(4)  &   86.4\\
16 &  1.210189(7)  &  0.826317(5)  &   256(4)   &   0.25(1)   &   23.4\\
17 &  1.228020(3)  &  0.814319(2)  &  1755(9)   &  -0.298(5)  &   62.4\\
18 &  1.24544(1)   &  0.802928(9)  &   130(3)   &   0.36(2)   &   13.4\\
19 &  1.259035(6)  &  0.794259(4)  &   464(5)   &   0.14(1)   &   32.3\\
20 &  1.27561(2)   &  0.78394(1)   &    60(2)   &   0.44(3)   &    9.8\\
21 &  1.28944(3)   &  0.77553(2)   &    31(1)   &  -0.37(4)   &    6.7\\
22 &  1.30351(3)   &  0.76716(2)   &    39(2)   &  -0.16(4)   &    7.8\\
23 &  1.31591(2)   &  0.75993(1)   &    49(2)   &   0.34(4)   &    8.0\\
24 &  1.32771(1)   &  0.753179(8)  &   124(3)   &   0.48(2)   &   13.5\\
25 &  1.340909(7)  &  0.745763(4)  &   362(4)   &  -0.12(1)   &   27.5\\
26 &  1.35445(3)   &  0.73831(1)   &    47(2)   &  -0.40(4)   &    8.0\\
27 &  1.36744(3)   &  0.73129(2)   &    30(1)   &   0.13(4)   &    6.5\\
28 &  1.37939(3)   &  0.72496(2)   &    30(1)   &   0.44(4)   &    6.3\\
29 &  1.39011(2)   &  0.71937(1)   &    63(2)   &  -0.20(3)   &    9.5\\
30 &  1.40098(2)   &  0.71379(1)   &    83(2)   &   0.29(3)   &   10.8\\
31 &  1.41017(2)   &  0.709133(9)  &   109(3)   &   0.42(2)   &   13.0\\
32 &  1.42162(3)   &  0.70342(2)   &    30(1)   &   0.38(4)   &    6.6\\
33 &  1.43338(4)   &  0.69765(2)   &    22(1)   &   0.08(5)   &    5.5\\
34 &  1.44328(5)   &  0.69287(2)   &    15.0(9) &   0.04(6)   &    4.2\\
35 &  1.45261(4)   &  0.68842(2)   &    23(1)   &   0.44(5)   &    5.8\\
36 &  1.46046(4)   &  0.68472(2)   &    26(1)   &  -0.15(5)   &    4.7\\
\tableline
\end{tabular}
\end{center}
\end{table*}


\section{Discussion}\label{conclusions}

We have identified the clearest and longest period series of consecutive high-order gravity modes in an SPB star to date. We see clear effects of rotation in the tilt of the period spacing series. Similar effects have also been found in $\gamma$\,Dora\-dus stars \citep{2014arXiv1411.1883B,2015A&A...574A..17V,2015arXiv150402119V}. The series here is not only tilted, but also shows periodic oscillations -- with an unprecedented clarity -- which carry information about the size of the convective core (the age of the star) and the mixing processes inside the star. Unlike for KIC\,10526294, which turned out to be too young to show these oscillations, or the first two detections where the period spacing series were too short to be conclusive, future detailed seismic modelling of this star will be able to put stringent constraints on the age, the core-overshoot parameter and diffusive chemical mixing properties of the star.

A compatibility-check based on the expected period spacing behaviour relying on Ledoux splitting for high-order g-modes \citep[see, e.g., Chapter 3 in][]{2010aste.book.....A} is shown in the bottom panel of Figure\,\ref{fig4}. The overall trend in the observed period series (both the tilt and the periodic deviations from it) can be reproduced from an appropriate stellar model with $\mathcal{M}=3.3\,\mathcal{M}_{\Sun}$, $X=0.71$, $Z=0.02$, central hydrogen fraction $X_\mathrm{c}=0.44$, and core overshoot $\alpha_\mathrm{ov}=0.3$, using $\ell=1,m=1$ modes assuming a rotation rate that is compatible with the measured $v \sin i$. The rotation rate might place the observed modes in the gravito-inertial regime, where the perturbative treatment of rotation is not valid anymore \citep{2010A&A...518A..30B}. Therefore, future in-depth modelling will have to take this into account after derivation of $\Omega(r)$, and also explain why the retrograde sectoral and the zonal dipole modes are not visible. It is noteworthy that \citet{2003MNRAS.343..125T} already provided quantitative arguments that favor the visibility of prograde dipole sectoral modes, completely in agreement with our observational results. Once we have a good seismic model of the star, we also plan to investigate the physical nature of the linear combination frequencies in terms of non-linear response of the stellar flux to one or a few high-amplitude modes \citep{1996MNRAS.281..696G} or due to resonant mode coupling \citep{1997A&A...321..159B}. The latter predicts specific resonances to occur among the various heat-driven modes. In line with such theoretical predictions, we can then check if there is more \textit{a posteriori} information in the frequency spectrum beyond the detected $\ell=1$ series. Such further interpretation can only be done after the dominant dipole series is adequately modelled.

This star has the potential of becoming the seismic \textit{Rosetta Stone} of SPB stars. While ground based mode identification efforts can at best only deliver constraints on a few oscillation frequencies \citep[e.g.,][]{2000A&A...362..189C,2015MNRAS.446.1438D}, the unambiguously detected series of 36 modes carry the seismic information needed to understand the physical processes at play inside SPB stars better. For the near future, we need to find more SPBs with such series, and we plan to invest in their in-depth seismic modelling.


\acknowledgments

The research leading to these results was based on funding from the Fund for Scientific Research of Flanders (FWO) under grant agreement G.0B69.13., as well as from the Belgian Science Policy Office (Belspo, C90309: CoRoT Data Exploitation), and from the Research Council of KU Leuven under grant GOA/2013/012, Belgium. This research was also supported in part by the National Science Foundation of the United States under Grant No.~NSF PHY11-25915. The results are based on observations made with the Mercator telescope, operated by the Flemish Community  on the island of La Palma at the Spanish Observatorio del Roque de los Muchachos of the Instituto de Astrof\'{i}sica de Canarias. These observations were obtained with the \textsc{Hermes} spectrograph, which is supported by the Fund for Scientific Research of Flanders (FWO), Belgium, the Research Council of KU Leuven, Belgium, the Fonds National Recherches Scientific (FNRS), Belgium, the Royal Observatory of Belgium, the Observatoire de Gen\`{e}ve, Switzerland and the Th\"{u}ringer Landessternwarte Tautenburg, Germany. Funding for the \textit{Kepler} mission is provided by the NASA Science Mission directorate. Some of the data presented in this paper were obtained from the Multimission Archive at the Space Telescope Science Institute (MAST). STScI is operated by the Association of Universities for Research in Astronomy, Inc., under NASA contract NAS5-26555. Support for MAST for non-HST data is provided by the NASA Office of Space Science via grant NNX09AF08G and by other grants and contracts. PIP and CA thank the staff of the Kavli Institute for Theoretical Physics at the University of California at Santa Barbara for their kind hospitality during their stay in February 2015.

{\it Facilities:} \facility{Mercator1.2m}.


\begin{thebibliography}{}
\expandafter\ifx\csname natexlab\endcsname\relax\def\natexlab#1{#1}\fi

\bibitem[{{Aerts}(2015)}]{2015IAUS..307..154A}
{Aerts}, C. 2015, in IAU Symposium, Vol. 307, IAU Symposium, 154--164

\bibitem[{{Aerts} {et~al.}(2010){Aerts}, {Christensen-Dalsgaard}, \&
  {Kurtz}}]{2010aste.book.....A}
{Aerts}, C., {Christensen-Dalsgaard}, J., \& {Kurtz}, D.~W. 2010,
  {Asteroseismology} ({Astronomy and Astrophsyics Library, Springer Berlin
  Heidelberg})

\bibitem[{{Aerts} {et~al.}(2014){Aerts}, {Sim{\'o}n-D{\'{\i}}az}, {Groot}, \&
  {Degroote}}]{2014A&A...569A.118A}
{Aerts}, C., {Sim{\'o}n-D{\'{\i}}az}, S., {Groot}, P.~J., \& {Degroote}, P.
  2014, \aap, 569, A118

\bibitem[{{Asplund} {et~al.}(2005){Asplund}, {Grevesse}, \&
  {Sauval}}]{2005ASPC..336...25A}
{Asplund}, M., {Grevesse}, N., \& {Sauval}, A.~J. 2005, in Astronomical Society
  of the Pacific Conference Series, Vol. 336, Cosmic Abundances as Records of
  Stellar Evolution and Nucleosynthesis, ed. {T.~G.~Barnes III \& F.~N.~Bash},
  25

\bibitem[{{Ballot} {et~al.}(2010){Ballot}, {Ligni{\`e}res}, {Reese}, \&
  {Rieutord}}]{2010A&A...518A..30B}
{Ballot}, J., {Ligni{\`e}res}, F., {Reese}, D.~R., \& {Rieutord}, M. 2010,
  \aap, 518, A30

\bibitem[{{Bedding} {et~al.}(2014){Bedding}, {Murphy}, {Colman}, \&
  {Kurtz}}]{2014arXiv1411.1883B}
{Bedding}, T.~R., {Murphy}, S.~J., {Colman}, I.~L., \& {Kurtz}, D.~W. 2014,
  ArXiv e-prints, arXiv:1411.1883

\bibitem[{{Borucki} {et~al.}(2010){Borucki}, {Koch}, {Basri}, {Batalha},
  {Brown}, {Caldwell}, {Caldwell}, {Christensen-Dalsgaard}, {Cochran},
  {DeVore}, {Dunham}, {Dupree}, {Gautier}, {Geary}, {Gilliland}, {Gould},
  {Howell}, {Jenkins}, {Kondo}, {Latham}, {Marcy}, {Meibom}, {Kjeldsen},
  {Lissauer}, {Monet}, {Morrison}, {Sasselov}, {Tarter}, {Boss}, {Brownlee},
  {Owen}, {Buzasi}, {Charbonneau}, {Doyle}, {Fortney}, {Ford}, {Holman},
  {Seager}, {Steffen}, {Welsh}, {Rowe}, {Anderson}, {Buchhave}, {Ciardi},
  {Walkowicz}, {Sherry}, {Horch}, {Isaacson}, {Everett}, {Fischer}, {Torres},
  {Johnson}, {Endl}, {MacQueen}, {Bryson}, {Dotson}, {Haas}, {Kolodziejczak},
  {Van Cleve}, {Chandrasekaran}, {Twicken}, {Quintana}, {Clarke}, {Allen},
  {Li}, {Wu}, {Tenenbaum}, {Verner}, {Bruhweiler}, {Barnes}, \&
  {Prsa}}]{2010Sci...327..977B}
{Borucki}, W.~J., {Koch}, D., {Basri}, G., {et~al.} 2010, Science, 327, 977

\bibitem[{{Breger} {et~al.}(2011){Breger}, {Balona}, {Lenz}, {Hollek}, {Kurtz},
  {Catanzaro}, {Marconi}, {Pamyatnykh}, {Smalley}, {Su{\'a}rez}, {Szabo},
  {Uytterhoeven}, {Ripepi}, {Christensen-Dalsgaard}, {Kjeldsen}, {Fanelli},
  {Ibrahim}, \& {Uddin}}]{2011MNRAS.414.1721B}
{Breger}, M., {Balona}, L., {Lenz}, P., {et~al.} 2011, \mnras, 414, 1721

\bibitem[{{Buchler} {et~al.}(1997){Buchler}, {Goupil}, \&
  {Hansen}}]{1997A&A...321..159B}
{Buchler}, J.~R., {Goupil}, M.-J., \& {Hansen}, C.~J. 1997, \aap, 321, 159

\bibitem[{{Chapellier} {et~al.}(2000){Chapellier}, {Mathias}, {Le Contel},
  {Garrido}, {Le Contel}, \& {Valtier}}]{2000A&A...362..189C}
{Chapellier}, E., {Mathias}, P., {Le Contel}, J.-M., {et~al.} 2000, \aap, 362,
  189

\bibitem[{{Daszy{\'n}ska-Daszkiewicz}
  {et~al.}(2015){Daszy{\'n}ska-Daszkiewicz}, {Dziembowski}, {Jerzykiewicz}, \&
  {Handler}}]{2015MNRAS.446.1438D}
{Daszy{\'n}ska-Daszkiewicz}, J., {Dziembowski}, W.~A., {Jerzykiewicz}, M., \&
  {Handler}, G. 2015, \mnras, 446, 1438

\bibitem[{{De Cat} {et~al.}(2005){De Cat}, {Briquet},
  {Daszy{\'n}ska-Daszkiewicz}, {Dupret}, {De Ridder}, {Scuflaire}, \&
  {Aerts}}]{2005A&A...432.1013D}
{De Cat}, P., {Briquet}, M., {Daszy{\'n}ska-Daszkiewicz}, J., {et~al.} 2005,
  \aap, 432, 1013

\bibitem[{{Degroote} {et~al.}(2010){Degroote}, {Aerts}, {Baglin}, {Miglio},
  {Briquet}, {Noels}, {Niemczura}, {Montalban}, {Bloemen}, {Oreiro}, {Vu{\v
  c}kovi{\'c}}, {Smolders}, {Auvergne}, {Baudin}, {Catala}, \&
  {Michel}}]{2010Natur.464..259D}
{Degroote}, P., {Aerts}, C., {Baglin}, A., {et~al.} 2010, \nat, 464, 259

\bibitem[{{Dolez} {et~al.}(2006){Dolez}, {Vauclair}, {Kleinman}, {Chevreton},
  {Fu}, {Solheim}, {Gonz{\'a}lez Perez}, {Ulla}, {Fraga}, {Kanaan}, {Reed},
  {Kawaler}, {O'Brien}, {Metcalfe}, {Nather}, {Sanwal}, {Klumpe}, {Mukadam},
  {Wood}, {Ahrens}, {Silvestri}, {Sullivan}, {Sullivan}, {Jiang}, {Xu},
  {Ashoka}, {Leibowitz}, {Ibbetson}, {Ofek}, {Kilkenny}, {Mei{\v s}tas},
  {Alisauskas}, {Janulis}, {Kalytis}, {Moskalik}, {Zola}, {Krzesinski},
  {Ogloza}, {Handler}, {Silvotti}, \& {Bernabei}}]{2006A&A...446..237D}
{Dolez}, N., {Vauclair}, G., {Kleinman}, S.~J., {et~al.} 2006, \aap, 446, 237

\bibitem[{{Dziembowski}(1977)}]{1977AcA....27...95D}
{Dziembowski}, W. 1977, \actaa, 27, 95

\bibitem[{{Dziembowski} {et~al.}(1993){Dziembowski}, {Moskalik}, \&
  {Pamyatnykh}}]{1993MNRAS.265..588D}
{Dziembowski}, W.~A., {Moskalik}, P., \& {Pamyatnykh}, A.~A. 1993, \mnras, 265,
  588

\bibitem[{{Garrido} \& {Rodriguez}(1996)}]{1996MNRAS.281..696G}
{Garrido}, R., \& {Rodriguez}, E. 1996, \mnras, 281, 696

\bibitem[{{Gautschy} \& {Saio}(1993)}]{1993MNRAS.262..213G}
{Gautschy}, A., \& {Saio}, H. 1993, \mnras, 262, 213

\bibitem[{{Gilliland} {et~al.}(2010){Gilliland}, {Brown},
  {Christensen-Dalsgaard}, {Kjeldsen}, {Aerts}, {Appourchaux}, {Basu},
  {Bedding}, {Chaplin}, {Cunha}, {De Cat}, {De Ridder}, {Guzik}, {Handler},
  {Kawaler}, {Kiss}, {Kolenberg}, {Kurtz}, {Metcalfe}, {Monteiro}, {Szab{\'o}},
  {Arentoft}, {Balona}, {Debosscher}, {Elsworth}, {Quirion}, {Stello},
  {Su{\'a}rez}, {Borucki}, {Jenkins}, {Koch}, {Kondo}, {Latham}, {Rowe}, \&
  {Steffen}}]{2010PASP..122..131G}
{Gilliland}, R.~L., {Brown}, T.~M., {Christensen-Dalsgaard}, J., {et~al.} 2010,
  \pasp, 122, 131

\bibitem[{{Hansen} {et~al.}(1985){Hansen}, {Winget}, \&
  {Kawaler}}]{1985ApJ...297..544H}
{Hansen}, C.~J., {Winget}, D.~E., \& {Kawaler}, S.~D. 1985, \apj, 297, 544

\bibitem[{{Lehmann} {et~al.}(2011){Lehmann}, {Tkachenko}, {Semaan},
  {Guti{\'e}rrez-Soto}, {Smalley}, {Briquet}, {Shulyak}, {Tsymbal}, \& {De
  Cat}}]{2011A&A...526A.124L}
{Lehmann}, H., {Tkachenko}, A., {Semaan}, T., {et~al.} 2011, \aap, 526, A124

\bibitem[{{Miglio} {et~al.}(2008){Miglio}, {Montalb{\'a}n}, {Noels}, \&
  {Eggenberger}}]{2008MNRAS.386.1487M}
{Miglio}, A., {Montalb{\'a}n}, J., {Noels}, A., \& {Eggenberger}, P. 2008,
  \mnras, 386, 1487

\bibitem[{{Nieva} \& {Przybilla}(2012)}]{2012A&A...539A.143N}
{Nieva}, M.-F., \& {Przybilla}, N. 2012, \aap, 539, A143

\bibitem[{{Pamyatnykh}(1999)}]{1999AcA....49..119P}
{Pamyatnykh}, A.~A. 1999, \actaa, 49, 119

\bibitem[{{P{\'a}pics}(2012)}]{2012AN....333.1053P}
{P{\'a}pics}, P.~I. 2012, Astronomische Nachrichten, 333, 1053

\bibitem[{{P{\'a}pics} {et~al.}(2014){P{\'a}pics}, {Moravveji}, {Aerts},
  {Tkachenko}, {Triana}, {Bloemen}, \& {Southworth}}]{2014A&A...570A...8P}
{P{\'a}pics}, P.~I., {Moravveji}, E., {Aerts}, C., {et~al.} 2014, \aap, 570, A8

\bibitem[{{P{\'a}pics} {et~al.}(2012){P{\'a}pics}, {Briquet}, {Baglin},
  {Poretti}, {Aerts}, {Degroote}, {Tkachenko}, {Morel}, {Zima}, {Niemczura},
  {Rainer}, {Hareter}, {Baudin}, {Catala}, {Michel}, {Samadi}, \&
  {Auvergne}}]{2012A&A...542A..55P}
{P{\'a}pics}, P.~I., {Briquet}, M., {Baglin}, A., {et~al.} 2012, \aap, 542, A55

\bibitem[{{P{\'a}pics} {et~al.}(2013){P{\'a}pics}, {Tkachenko}, {Aerts},
  {Briquet}, {Marcos-Arenal}, {Beck}, {Uytterhoeven}, {Trivi{\~n}o Hage},
  {Southworth}, {Clubb}, {Bloemen}, {Degroote}, {Jackiewicz}, {McKeever}, {Van
  Winckel}, {Niemczura}, {Gameiro}, \& {Debosscher}}]{2013A&A...553A.127P}
{P{\'a}pics}, P.~I., {Tkachenko}, A., {Aerts}, C., {et~al.} 2013, \aap, 553,
  A127

\bibitem[{{Puls} {et~al.}(2005){Puls}, {Urbaneja}, {Venero}, {Repolust},
  {Springmann}, {Jokuthy}, \& {Mokiem}}]{2005A&A...435..669P}
{Puls}, J., {Urbaneja}, M.~A., {Venero}, R., {et~al.} 2005, \aap, 435, 669

\bibitem[{{Raskin} {et~al.}(2011){Raskin}, {van Winckel}, {Hensberge},
  {Jorissen}, {Lehmann}, {Waelkens}, {Avila}, {de Cuyper}, {Degroote},
  {Dubosson}, {Dumortier}, {Fr{\'e}mat}, {Laux}, {Michaud}, {Morren}, {Perez
  Padilla}, {Pessemier}, {Prins}, {Smolders}, {van Eck}, \&
  {Winkler}}]{2011A&A...526A..69R}
{Raskin}, G., {van Winckel}, H., {Hensberge}, H., {et~al.} 2011, \aap, 526, A69

\bibitem[{{Schmidt-Kaler}(1982)}]{1982SchmidtKalerBook}
{Schmidt-Kaler}, T. 1982, {Landolt-B{\"o}rnstein}, ed. {K.~Schaifers \&
  H.~H.~Vogt}, Vol.~2b ({Springer--Verlag})

\bibitem[{{Seaton}(2005)}]{2005MNRAS.362L...1S}
{Seaton}, M.~J. 2005, \mnras, 362, L1

\bibitem[{{Tkachenko} {et~al.}(2012){Tkachenko}, {Lehmann}, {Smalley},
  {Debosscher}, \& {Aerts}}]{2012MNRAS.422.2960T}
{Tkachenko}, A., {Lehmann}, H., {Smalley}, B., {Debosscher}, J., \& {Aerts}, C.
  2012, \mnras, 422, 2960

\bibitem[{{Townsend}(2000)}]{2000MNRAS.318....1T}
{Townsend}, R.~H.~D. 2000, \mnras, 318, 1

\bibitem[{{Townsend}(2003)}]{2003MNRAS.343..125T}
---. 2003, \mnras, 343, 125

\bibitem[{{Triana} {et~al.}(2015){Triana}, {Moravveji}, {P\'apics}, {Aerts},
  {Kawaler}, \& {Christensen-Dalsgaard}}]{Triana2015}
{Triana}, S.~A., {Moravveji}, E., {P\'apics}, P.~I., {et~al.} 2015, \apj,
  submitted

\bibitem[{{Van Reeth} {et~al.}(2015{\natexlab{a}}){Van Reeth}, {Tkachenko},
  {Aerts}, {P{\'a}pics}, {Degroote}, {Debosscher}, {Zwintz}, {Bloemen}, {De
  Smedt}, {Hrudkova}, {Raskin}, \& {Van Winckel}}]{2015A&A...574A..17V}
{Van Reeth}, T., {Tkachenko}, A., {Aerts}, C., {et~al.} 2015{\natexlab{a}},
  \aap, 574, A17

\bibitem[{{Van Reeth} {et~al.}(2015{\natexlab{b}}){Van Reeth}, {Tkachenko},
  {Aerts}, {Papics}, {Triana}, {Zwintz}, {Degroote}, {Debosscher}, {Bloemen},
  {Schmid}, {De Smedt}, {Fremat}, {Fuentes}, {Homan}, {Hrudkova},
  {Karjalainen}, {Lombaert}, {Nemeth}, {Oestensen}, {Van De Steene}, {Vos},
  {Raskin}, \& {Van Winckel}}]{2015arXiv150402119V}
---. 2015{\natexlab{b}}, ArXiv e-prints, arXiv:1504.02119

\end{thebibliography}

\end{document}